# Modulational instability and solitary waves in polariton topological insulators

YAROSLAV V. KARTASHOV[1,2,3,*] AND DMITRY V. SKRYABIN[3,4]

[1]ICFO-Institut de Ciencies Fotoniques, The Barcelona Institute of Science and Technology, 08860 Castelldefels (Barcelona), Spain
[2]Institute of Spectroscopy, Russian Academy of Sciences, Troitsk, Moscow Region, 142190, Russia
[3]Department of Physics, University of Bath, BA2 7AY, Bath, United Kingdom
[4]ITMO University, Kronverksky Avenue 49, St. Petersburg 197101, Russia
*Corresponding author: Yaroslav.Kartashov@icfo.es

**Optical microcavities supporting exciton-polariton quasi-particles offer one of the most powerful platforms for investigation of rapidly developing area of topological photonics in general, and of photonic topological insulators in particular. Energy bands of the microcavity polariton graphene are readily controlled by magnetic field and influenced by the spin-orbit coupling effects, a combination leading to formation of linear unidirectional edge states in polariton topological insulators as predicted very recently. In this work we depart from the linear limit of non-interacting polaritons and predict instabilities of the nonlinear topological edge states resulting in formation of the localized topological quasi-solitons, which are exceptionally robust and immune to backscattering wavepackets propagating along the graphene lattice edge. Our results provide a background for experimental studies of nonlinear polariton topological insulators and can influence other subareas of photonics and condensed matter physics, where nonlinearities and spin-orbit effects are often important and utilized for applications.**

## 1. Introduction

Topological insulators and topologically protected edge states attract nowadays unprecedented attention in diverse areas of science, including solid-state physics, acoustics, matter waves, graphene-based applications and photonics, see, e.g., [1,2] for recent reviews. Topological insulators are characterized by the presence of the complete band-gap in the bulk of the material, like in a usual insulator, while at the same time they admit in-gap edge states propagating at the surface, where conduction of electrons becomes possible in the presence of magnetic field. These edge states are generally allowed if materials placed in contact have bulk Hamiltonians characterized by different topological invariants – Chern numbers. Apart from their conductive properties, edge states in topological insulators are immune to backscattering through their topological protection [1,2]. Topological edge states were shown to exist in HgTe quantum wells, BiSb alloys, and some other materials, where their unidirectional character and immunity to backscattering were also confirmed [1,2]. Spin-orbit interactions for electrons is the key phenomenon underpinning existence of the topological insulator phase, whose physics is closely linked with quantum Hall effect and integer Hall conductance [1,2].

Electromagnetic topological edge states have also been under intense investigation [3]. They were predicted and observed in gyromagnetic photonic crystals with pronounced Faraday effect in microwave range [4,5], in arrays of coupled resonators [6,7], and in metamaterial superlattices [8]. One of the most spectacular realizations of unidirectional edge states at optical frequencies was reported in honeycomb arrays of helical waveguides [9].

Electronic and photonic edge states mentioned above are purely linear. Though nonlinearities associated with either optical transitions or with inter-particle interactions are inherent in the majority of optical systems, the investigation of their impact on non-topological and topological edge states is in its infancy at this moment. Thus, *non-topological* edge states in photonic graphene have been recently studied in [10], the edge states were shown to exist in the presence of nonlinearity, while an attempt to observe soliton effects gave initial localization subsequently accompanied by noticeable radiation into the bulk, indicating coupling to the extended modes of the lattice. Nonlinearity was shown to strongly affect transmission and reflection of edge modes at the corners of graphene-like photonic structures [11]. Photonic graphene stripes of small width (ribbons), where the edge effects are mixed with the bulk dispersion, were considered in [12] and various *non-topological* nonlinear localized states were found, which bear more from the solitons in the bulk of the lattice.

Even more rare results on nonlinear *topological* states that were obtained so far are connected with scalar optical models describing evolution of excitations in arrays of twisted waveguides. Bulk nonlinear modes (i.e. modes located in the depth of periodic structure and not on its surface) of topological insulator made of helical waveguides were obtained in [13]. Traveling topological states in helical arrays were constructed in [14], but only in the frames of simplified discrete model. Dynamical excitation of their very well localized continuous counterparts considered in the unpublished Ref. [15] illustrates that longitudinal oscillations of waveguides introduce strong radiative losses leading to notable reduction of peak power already after traversing of ten sites of the structure. We should also mention here a paper linking topology and nonlinearity in a one dimensional dimer chain [16].

Phenomenology of topological insulators and edge states was recently transferred to a rapidly developing domain of exciton-polaritons in microcavities [17,18]. Main advantages of polaritons include sufficiently strong spin-orbit coupling originating in the cavity induced TE-TM splitting of the polariton energy levels [19,20], established technology of the microcavity structuring into arbitrary lattice potentials [19,21], and very strong nonlinear interactions of polaritons through their excitonic component. The latter was used for recent demonstrations of superfluidity [22,23], generation of dark quasi-solitons and vortices [24-27], bright spatial and temporal solitons [28-30], and other effects. The observed polariton effects with linear and nonlinear lattice potentials include one- [31] and two-dimensional [32,33] gap polariton solitons, visualization of Dirac cones [34] and flat bands [35], and visualization of non-topological edge states [21]. Recently, it has been shown theoretically that attractive nonlinear interaction between polaritons with opposite spins can compensate and exceed Zeeman energy shifts due to magnetic field and thereby lead to the inversion of the propagation direction of the edge states [36]. Apart from this result the nonlinear effects with topological polariton edge states remain unexplored.

Thus, to the best of our knowledge, robust (long-living) nonlinear topological edge states confined in the direction parallel to the surface were never demonstrated in the frames of continuous physical models and in real-world systems, where unidirectionality and topological protection are provided by physical effects different from basic temporal variation of the underlying potential that always introduces undesirable losses. Moreover, compact topologically protected nonlinear edge states were never addressed in nowadays rapidly developing and open for experimental exploration spinor systems, such as spin-orbit coupled polariton and Bose-Einstein condensates, that may feature much richer dynamics and provide principally new tools for control of the state of the system in comparison with conventional scalar settings. It should be stressed that realization of topological insulators supporting surface transport of *localized nonspreading excitations* over considerable time intervals is a problem of fundamental physical importance, whose solution may pave the way to a number of practical applications.

We show here one such system is represented by the interacting polariton topological insulator. We use continuous model for spin-orbit coupled polariton condensate in honeycomb arrays accounting for spin-dependent interactions and Zeeman splitting in the external magnetic field to demonstrate a variety of new nonlinear topologically protected edge states and to perform their rigorous stability analysis, for the first time for this class of topologically protected nonlinear modes. Extended periodic edge states are found in the exact form as truly stationary nonlinear solutions bifurcating from linear periodic edge states. We show that their modulation instability results in splitting of the extended nonlinear edge states into sets of fully localized edge quasi-solitons travelling along the interface over very long time intervals without notable deformations. Such localized quasi-solitons appear to be very robust objects on any practical time scales. Even though they radiate as they move along the interface, the rate of radiation can be exceptionally small in the proper range of parameters ensuring that such states can traverse huge distances along the surface of the material without being destroyed or scattered. The approximate expression for the shape of quasi-solitons is derived.

## 2. Topological edge states in the linear regime

Polariton condensate in the lattice potential in the presence of the external magnetic field can be described by a system of coupled Gross-Pitaevskii equations for the spin-positive and spin-negative components $\psi_\pm$ of the spinor polariton wavefunction $\mathbf{\Psi} = (\psi_+, \psi_-)^\mathrm{T}$ [17,37]:

$$i\hbar \frac{\partial \psi_\pm}{\partial t} = -\frac{\hbar^2}{2m^*}\left(\frac{\partial^2}{\partial x^2} + \frac{\partial^2}{\partial y^2}\right)\psi_\pm + \frac{\beta\hbar^2}{m^*}\left(\frac{\partial}{\partial x} \mp i\frac{\partial}{\partial y}\right)^2 \psi_\mp + \quad (1)$$
$$[\varepsilon_0 \mathcal{R}(x,y) \pm \varepsilon_z]\psi_\pm + \varepsilon_0(|\psi_\pm|^2 + \sigma|\psi_\mp|^2)\psi_\pm.$$

Quasi-conservative nonlinear dynamics has been observed in several experiments with exciton polaritons, see, e.g. [19,24,30,31], and used in many theoretical studies, see, e.g. [17-19,37]. Following this trend we have also chosen to work here in the idealized conservative limit, since the very fact of existence of unidirectional edge states is not connected with presence of losses, so we do not include them to have most transparent picture of the phenomenon. Here the relation between the wavefunction components in the circular polarization basis and those associated with TE (subscript $y$) and TM (subscript $x$) polarizations is given by $\psi_\pm = (\psi_\mathrm{x} \mp i\psi_\mathrm{y})/2^{1/2}$. The $\beta$-term describes spin-orbit coupling originating from the TE-TM energy splitting of the cavity photons, which translates into slightly different effective masses $m_{x,y}$ of TM and TE polaritons: $\beta = (m_x - m_y)/4m^*$. Accounting for the spinor nature of the system and transforming into the basis of circular polarizations results in the spin-orbit term raised to the second power, see, e.g., [37]. Parameter $\varepsilon_z$ is the Zeeman energy splitting of spin + and spin – polaritons proportional to the externally applied magnetic field. $\varepsilon_0 \mathcal{R}(x,y)$ describes potential energy landscape in the microcavity [19,21,32,33]. In our case $\mathcal{R}(x,y)$ is a honeycomb lattice that is cut in the $x$ direction. The distance between the lattice sites is $a$, so that $\mathcal{R}(x,y) = \mathcal{R}(x, y + a3^{1/2})$. $\varepsilon_0$ is the scaling coefficient with the dimension of energy, while $\mathcal{R}$ is the dimensionless function. Amplitudes $\psi_\pm$ can be assumed dimensionless and scaled in a way that the nonlinear energy shift achieved for the unit polariton density equals $\varepsilon_0$. This scaling well reflects the physically realistic situation, when energies, associated with the lattice potential, Zeeman effect and nonlinearity have the same order of magnitude [17-19]. $\sigma = -0.05$ is the strength of the weak attractive interaction of polaritons with the opposite spins [38]. Local potential minima in $\mathcal{R}(x,y)$ are described by Gaussian functions $-pe^{-[(x-x_n)^2 + (y-y_n)^2]/d^2}$ with depth $p$ and width $d$.

We scale physical distance with $L = 3^{1/2}$ μm and hence all energy parameters are conveniently scaled with the characteristic energy $\varepsilon^* = \hbar^2/m^*L^2 \simeq 0.1$ meV for $m^* = 10^{-31}$ g. Without any restriction of generality, since $\mathcal{R}$ contains a factor $p$ and nonlinear terms can be scaled arbitrarily, we choose $\varepsilon_0 = \varepsilon^*$ and normalize physical time with $\hbar\varepsilon_0^{-1} \simeq 6$ ps. In what follows the potential depth is $p = 8$, corresponding to $0.8$ meV, the width of individual potential wells is $d = 0.5$ and separation between minima is $a = 1.4$ in dimensionless length units, that corresponds to $0.87$ μm and $2.42$ μm, respectively. In what follows the dimensionless Zeeman splitting parameter $\Omega = \varepsilon_z/\varepsilon_0$ is chosen to be 0.5 and the spin-orbit parameter $\beta = 0.3$ (unless stated otherwise), so that both of them are an order of magnitude less than the energy shift induced by the lattice. Note, that the parabolic approximation for the polariton energy momentum dependence dictated by the Gross-Pitaevskii approximation is well obeyed providing that all other energy shifts in our model are less than the width of the low polariton branch by a factor of 2-3 or more, which is well satisfied if we realistically assume that the latter is ~10 meV wide. With these normalizations the dimensionless version of Eq. (1) can be written as

$$i\frac{\partial \psi_\pm}{\partial t} = -\frac{1}{2}\left(\frac{\partial^2}{\partial x^2} + \frac{\partial^2}{\partial y^2}\right)\psi_\pm + \beta\left(\frac{\partial}{\partial x} \mp i\frac{\partial}{\partial y}\right)^2 \psi_\mp + \quad (2)$$
$$\mathcal{R}(x,y)\psi_\pm \pm \Omega\psi_\pm + (|\psi_\pm|^2 + \sigma|\psi_\mp|^2)\psi_\pm,$$

where we retained old notations for scaled coordinates $x, y$ and evolution time $t$.

For these parameters and for $\beta, \Omega \neq 0$ the simplest linear mode of an isolated local minimum of the potential is characterized by the presence of vortex with topological charge 2 in $\psi_+$ component with ring-like density profile and by trivial constant phase of bell-shaped $\psi_-$ component. Inter-estingly, next mode carries vortex in the $\psi_-$ component, while $\psi_+$ component has trivial phase. The appearance of charge-2 vortex in one of the components is a consequence of spin-orbit coupling and it can be observed only for $\beta \neq 0$. If the potential has two well-separated minima, the above mentioned charge-2 vortices carried by one of the components and located around each potential minima split into two charge-1 vortices. When two potential minima become very close, one observes linear modes with complex phase distributions with more than 4 phase singularities in one of the components.

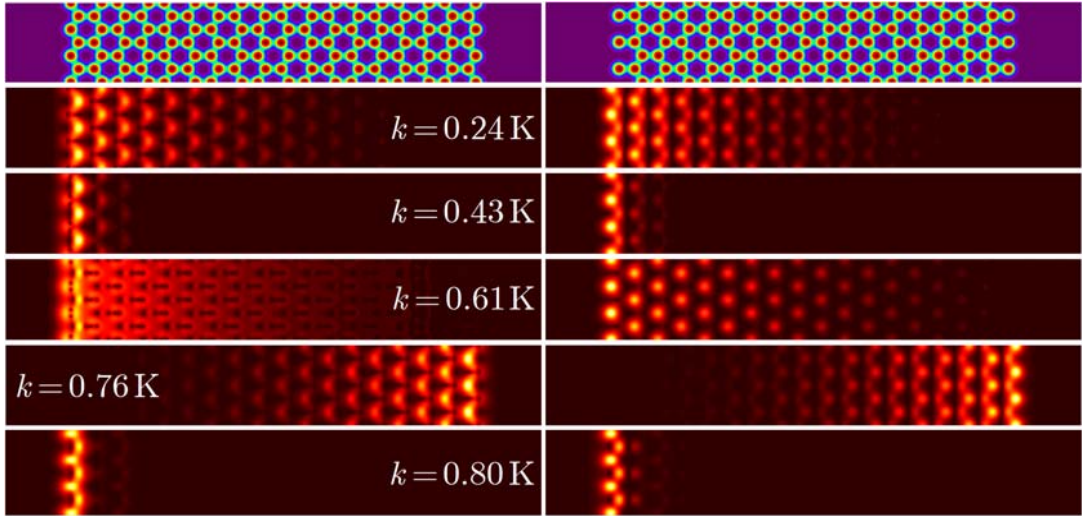

Fig. 1. (Color online) First row shows examples of the array with zigzag and bearded edges (three periods of the structure are shown in the $y$ direction). Second, third and fourth rows illustrate transformation of the edge state profiles corresponding to the red lines (top gap) in Figs. 2(c), 3(a) upon variation of Bloch momentum $k$. Fifth row shows the edge state on the other end of the lattice. Sixth row shows the edge state corresponding to the magenta line from the bottom gap. In all cases $x \in [-24, +24]$ window is shown, $\beta = 0.3$, $\Omega = 0.5$. Left and right columns show $|\psi_+|$ and $|\psi_-|$, respectively.

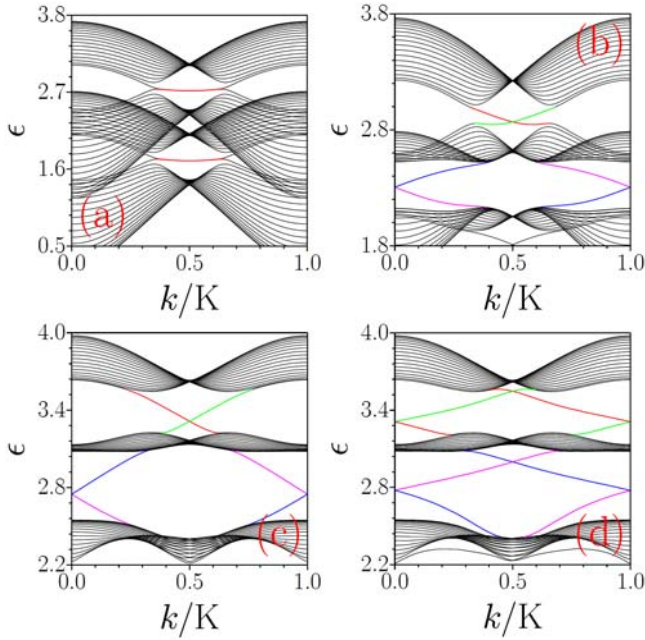

Fig. 2. (Color online) Energy-momentum diagrams $\varepsilon(k)$ obtained for the lattice with zigzag edges at $\beta = 0$ (a), $\beta = 0.15$ (b), $\beta = 0.3$ (c) and for the lattice with bearded edges at $\beta = 0.3$ (d). Black lines correspond to modes residing in the bulk of the lattice, while color lines indicate edge states. Topologically protected edge states in (b)-(d) with opposite slopes of $\varepsilon(k)$ belong to the opposite edges of the lattice. In all cases $\Omega = 0.5$.

There are three types of termination of extended honeycomb arrays corresponding to the zigzag (Fig. 1, top left), bearded (Fig. 1, top right), and armchair edges. The armchair cut usually does not support the edge states [9], so we do not consider it here. We first address the spectrum of linear modes that are periodic along the array edge and localized along the $x$-axis on both sides from the edge. These modes are the Bloch functions $\psi_\pm(x,y,t) = u_\pm(x,y)e^{iky+i\varepsilon t}$, where $u_\pm(x,y) = u_\pm(x, y+3^{1/2}a)$, $u_\pm|_{x\to\pm\infty} = 0$, $k$ is the Bloch momentum, $\varepsilon$ is the energy shift relative to the bottom of the polariton energy-momentum characteristic, and $3^{1/2}a$ is the $y$-period of the potential. The corresponding momentum giving the width of the Brillouin zone is $K = 2\pi/3^{1/2}a$. We calculated Bloch functions numerically using a unit cell containing 36 potential minima (top row of Fig. 1 depicts three such unit cells, i.e. three periods of potential along the $y$-axis). Representative spectra for the lattice with *zigzag* and *bearded* edges are shown in Fig. 2 in the form of the energy-momentum diagrams $\varepsilon(k)$ for different $\beta, \Omega$ values and for $k \in [0, K]$ interval, rather than more traditional $k \in [-K/2, +K/2]$. Due to spinor character of the model the spectrum consists of two groups of bands. At $\beta, \Omega = 0$ these two groups are fully degenerate, and correspond to the Bloch modes with $u_+ \neq 0$, $u_- = 0$ and $u_+ = 0$, $u_- \neq 0$. The inclusion of Zeeman splitting lifts this degeneracy and results in vertical splitting of two energy bands by $2\Omega$, clearly visible in Fig. 2(a) for zigzag edge. The spectrum in Fig. 2(a) shows two Dirac points at $k = \pm K/3$, where first and second bands in each group touch each other. These points are traces of Dirac points in the spectrum of bulk honeycomb lattice. Red branches in Fig. 2(a) correspond to the non-topological edge states appearing due to truncation of the array, while black branches correspond to modes concentrated in the bulk. In the bearded edge case the edge modes appear in the region $k \in [-K/3, +K/3]$ between two Dirac points, while for zigzag edge they appear outside this domain, at $k \in [K/3, 2K/3]$. It should be stressed that without spin-orbit coupling, at $\beta = 0$, the edge states are almost dispersionless with $|\partial^n \varepsilon/\partial k^n| \ll 1$. The flatness of the energy-momentum characteristic for the edge states becomes practically perfect for larger separations $a$ between the potential minima. Also, for $\beta=0$, the edge states residing at the opposite ends of the lattice have identical energies.

This picture qualitatively changes when spin-orbit coupling is accounted for, see Figs. 2(b), 2(c) obtained for zigzag edge. It leads to opening of the full gap between the first (upper most) and second bands, and eliminates Dirac points. The width of the gap increases with $\beta$. Most importantly, spin-orbit coupling removes degeneracy of the edge states, so that the states residing on the opposite sides of the array acquire opposite group velocities $\partial\varepsilon/\partial k$, associated unidirectionality, and topological protection features [17]. The degeneracy is removed and edge states become unidirectional only for nonzero values of $\beta$ (the effect is observable already at $\beta\sim0.01$). Thus, spin-orbit coupling is an absolutely necessary ingredient for observation of all effects associated with unidirectionality that are mentioned below. It is SO-coupling that leads to nonzero Chern number and topological character of unidirectional edge states, both linear ones and bifurcating from them nonlinear modes. Moreover, spin-orbit coupling leads to specific band folding [Fig. 2(b)] accompanied by opening of the additional lower energy gaps, where topologically protected edge states appear in the intervals of momenta roughly complimentary to the ones where the edge states of the primary gap exist. To the best of our knowledge, the counterparts of these states were not encountered in scalar (non-spinor) honeycomb lattice models with zigzag termination [21] and in the Floquet topological insulators formed by helical waveguides [9]. Red (residing in the top gap) and magenta (residing in the bottom gap) curves in Figs. 2(b),2(c) correspond to modes on the left edge of the lattice propagating upwards, while green (top gap) and blue (bottom gap) curves correspond to the downwards propagating modes on the right edge. Dispersion diagrams obtained for array with bearded edges also reveal presence of edge states in top and bottom gaps [Fig. 2(d)].

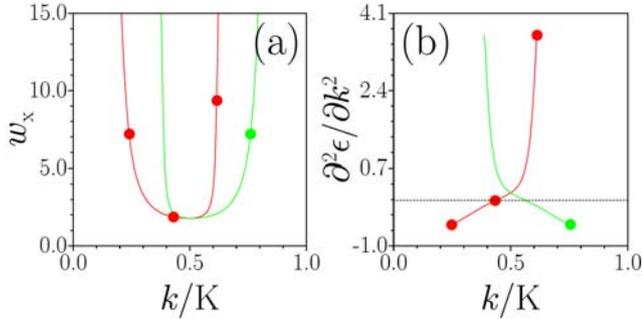

Fig. 3. (Color online) Width $w_x$ of linear edge states along the $x$ axis (a) and second-order dispersion coefficient $\partial^2\varepsilon/\partial k^2$ for branches associated with edge states (b) versus $k$ at $\beta=0.3$, $\Omega=0.5$. Red and green circles correspond to linear modes depicted in Fig. 1. Line colors correspond to colors used in Fig. 2(c) to denote different branches.

Examples of the profiles for the edge states in the top and bottom gaps are shown in Fig. 1 for the same parameters as used in Fig. 2(c) (zigzag edges). Edge states extend into the bulk of the lattice when momentum $k$ approaches the boundary of the existence domain (second and fourth rows), but are well localized for $k$ values returning energies close to the center of the gap (third row). The modes residing at the opposite edges, but having the same energies for different momenta are "mirror images" of each other (compare fifth and second rows). Edge states from the bottom gap associated with magenta line in Fig. 2(c) also can be well-localized (see sixth row in Fig. 1) provided that their energy is far from the gap edges. Note, that the $\varepsilon(k)$ plots for the topological edge states in our system are *not* antisymmetric with respect to the $k=K/2$ line, as in arrays of twisted waveguides [9]. The degree of asymmetry is controlled by the spin-orbit coupling. The asymmetry of the existence intervals with respect to the $k=K/2$ point and stronger localization around this point can be well seen in the dependence of the edge state width $w_x$ on the momentum [Fig. 3(a)]. The width is defined as $w_x=[2/(\chi_++\chi_-)]^{1/2}$, where

$$\chi_\pm = U_\pm^{-2}\int_{-\pi/K}^{+\pi/K}dy\int_{-\infty}^{\infty}|\psi_\pm|^4\,dx,$$
$$U_\pm = \int_{-\pi/K}^{+\pi/K}dy\int_{-\infty}^{\infty}|\psi_\pm|^2\,dx \qquad (3)$$

are the integral form-factors and the norms for the two spin components defined on one $y$-period of the structure, respectively.

### 3. Nonlinear topological edge states and their modulational instability

Solutions accounting for nonlinearity are sought in the same form as the linear ones $\psi_\pm(x,y,t)=u_\pm(x,y)e^{iky+i\mu t}$. Because self-repulsive nonlinearity in our model dominates over the weak counter-spin attraction, nonlinear solutions at a given $k$ exists only for $\mu<\varepsilon$, with the nonlinearity induced energy shift (chemical potential) $\mu$ becoming $\varepsilon$ in the linear limit. Nonlinear Bloch functions $u_\pm(x,y)$ satisfy

$$-\mu u_\pm = -\frac{1}{2}\left(\frac{\partial^2}{\partial x^2}+\frac{\partial^2}{\partial y^2}+2ik\frac{\partial}{\partial y}-k^2\right)u_\pm + \mathcal{R}(x,y)u_\pm +$$
$$\beta\left(\frac{\partial}{\partial x}\mp i\left(\frac{\partial}{\partial y}+ik\right)\right)^2 u_\mp \pm \Omega u_\pm + (|u_\pm|^2+\sigma|u_\mp|^2)u_\pm. \qquad (4)$$

We solved Eq. (4) for $u_\pm(x,y)$ numerically using Newton method in frequency domain, thus these solutions are obtained exactly as truly stationary modes of governing Eq. (1). Examples of the transverse profiles of these functions are shown in Fig. 4. We characterize nonlinear edge states by the dependence of their total norm $U=U_++U_-$ per $y$-period on $\mu$ (see Fig. 5). On the same figure we plot also the dependencies of the peak amplitudes $a_+=\max|\psi_+|$ and $a_-=\max|\psi_-|$ of two component vs $\mu$ as the nonlinear states bifurcate from the linear ones. Fig. 5 explicitly shows that the nonlinear edge states bifurcate backwards in $\mu$ from their linear counterparts. Since for the selected "upper" group of bands, the $\psi_-$ component in all linear modes has higher amplitude than the $\psi_+$ component, one has $a_->a_+$ also for nonlinear states. The situation will reverse if one changes direction of the applied magnetic field. $a_\pm$ vanish at the bifurcation point, indicating the thresholdless character of the unidirectional nonlinear edge states. The width of the existence domain in $\mu$ for nonlinear modes is determined by the difference between the energy of the linear edge state and the upper boundary of the band where the bulk modes reside for a given $k$. This means that the *localized* along the $x$-axis nonlinear edge states can have nonlinear energy shifts $\mu$ that are *smaller* than the lower edge of the *total* gap defined over all momenta $k$ [for instance at $k=K/2$ localized nonlinear modes can have $\mu$ values that are already within continuous band for $k=K/3$, see Fig. 2(c)]. When $\mu$ crosses the edge of the band for a given $k$ (shown by dashed lines in Fig. 5) the nonlinear mode loses localization due to coupling with the bulk modes. X-width of nonlinear modes monotonically increases toward the edge of the band [Fig. 5(c)]. Presence of multiple dark spots (zeroes) in $|\psi_+|$ distribution in Fig. 4 surrounded by brighter regions indicates existence of multiple vortices in $\psi_+$ component. This is a natural consequence of spin-orbit coupling in our model: if one component features local maximum around certain pillar, the coupling leads to appearance of charge-2 vortex in other component in this location. These vortices usually split into two charge-1 vortices due to perturbations introduced by neighboring pillars, so resulting phase distribution may be rather complex.

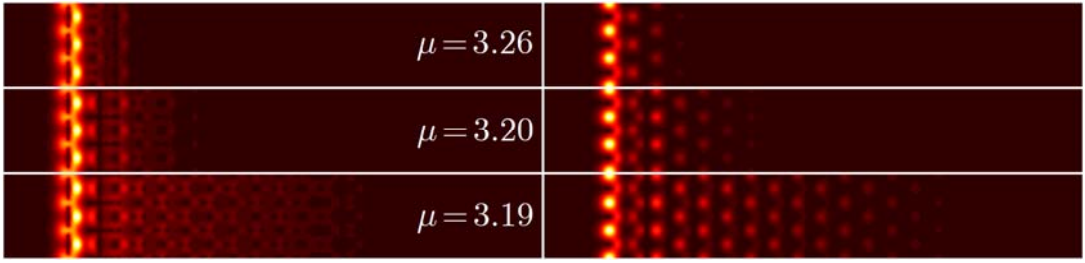

Fig. 4. (Color online) Transformation of the nonlinear edge state bifurcating from the red branch in Fig. 2(c) at $k=0.55\,\mathrm{K}$ with decrease of $\mu$. Left column shows $|\psi_+|$, right column $|\psi_-|$. In all cases $\beta=0.3$, $\Omega=0.5$.

Stability analysis of nonlinear edge states $u_\pm(x,y)$ was performed by perturbing them with small (1% in amplitude) broadband input noise and direct modeling of their subsequent evolution up to very large times (usually up to $t>10^4$). Nonlinear edge states were found to be unstable (recall that in the presence of periodic potentials modulation instability is possible even in the medium with defocusing nonlinearity [39,40]). The instability is particularly pronounced close to the edge of the band and is strongly suppressed when one approaches a point, where the nonlinear edge states bifurcate from the linear spectrum. Only low-frequency perturbations with huge $y$-periods substantially exceeding separation between pillars can destabilize such modes and their growth rates are so small that the instability manifests itself at times $t>10^3$ far exceeding lifetime of polaritons, so such modes will appear as stable ones in experiments. Modulation instability bandwidth (i.e. the range of frequency of modulations along the $y$ axis that can seed instability) becomes smaller with increase of $\mu$. Such bandwidth can be calculated from Eq. (2) using initial conditions $\psi_\pm(t=0)=u_\pm(x,y)[1+\nu\cos(\kappa y)]\exp(iky)$, where $\nu$ and $\kappa$ are the amplitude and frequency of small perturbations. Such perturbations experience clear exponential growth at the initial stage of instability development as long as $\kappa$ is within modulation instability band, that allows to determine instability growth rate $\delta$ as a function of $\kappa$. Fig. 5(d) shows representative $\delta(\kappa)$ dependence that reveals the finite instability bandwidth.

(a) and $k=0.55\,\mathrm{K}$ (b) bifurcating from red branch in Fig. 2(c) at $\mu=\varepsilon$. (c) X-width of nonlinear edge state versus $\mu$ at $k=0.55\,\mathrm{K}$. Circles in (c) correspond to the states shown in Fig. 4. Dashed lines indicate the edge of the band for corresponding $k$ values. (d) Perturbation growth rate for nonlinear edge state with $\mu=3.21$ versus frequency of small modulation $\kappa$. In all cases $\beta=0.3$, $\Omega=0.5$.

Typical dynamics of instability development stimulated by broadband noise is shown in Fig. 6. Instability initially leads to pronounced modulation of the edge state in the $y$-direction. As time goes this modulation is becoming accompanied by radiation into the depth of the array. However, instead of decay and disintegration, the edge state breaks up into a set of weakly radiating, but fully localized solitons (see the pattern at $t=960$ in Fig. 6). This suggests that predominantly repulsive nonlinearity and $\partial^2\varepsilon/\partial k^2>0$ can give rise to *unidirectional bright quasi-solitons*. These can be constructed as in-gap moving topological solitons, existing only at the edge of the array, and bifurcating from linear topological edge states.

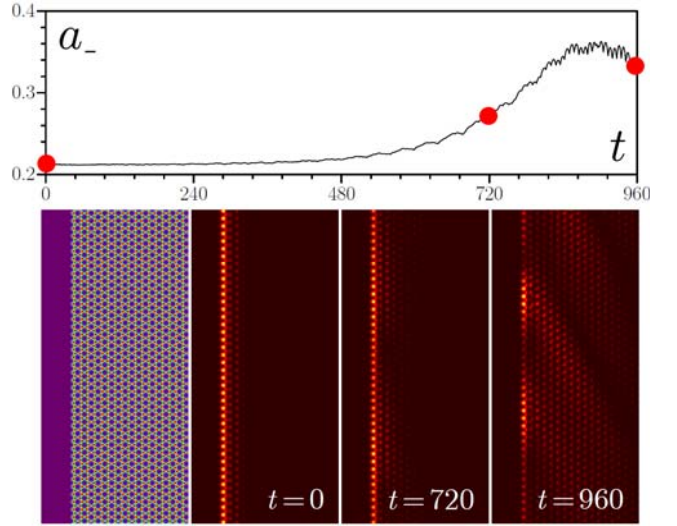

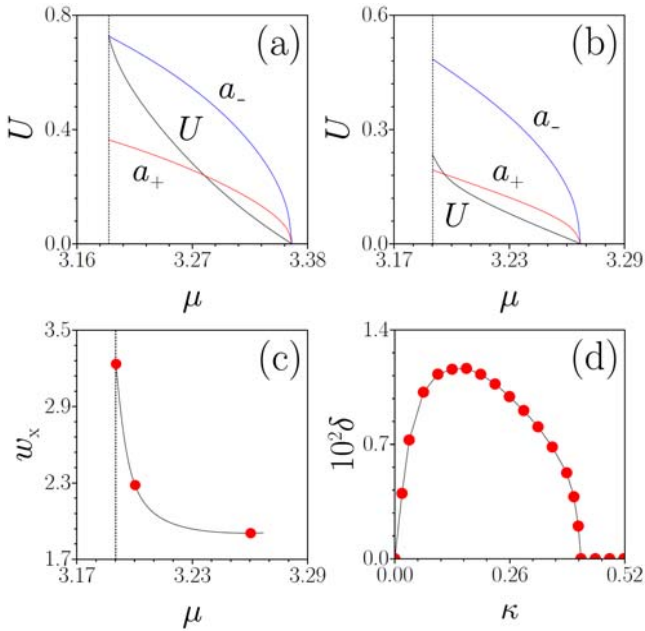

Fig. 5. (Color online) Norm $U$ per period and peak amplitudes $a_+,a_-$ of the spin components versus $\mu$ for nonlinear edge states with $k=0.45\,\mathrm{K}$

Fig. 6. (Color online) Instability development for the nonlinear edge state with $\mu=3.25$, $k=0.55\,\mathrm{K}$ stimulated by small input noise. First row shows peak amplitude of $\psi_-$ component versus time, while second row shows array and $|\psi_-|$ distributions in different moments of time corresponding to red circles. Here $\beta=0.3$, $\Omega=0.5$.

## 4. Edge quasi-solitons

To develop more regular approach to the question of quasi-soliton existence we rewrite Eq. (2) in the form $i\partial\Psi/\partial t=\mathcal{L}\Psi+\mathcal{N}\Psi$, where $\Psi=(\psi_+,\psi_-)^\mathrm{T}$, the operator $\mathcal{L}$ accumulates all linear terms, while op-erator $\mathcal{N}$ accounts for nonlinear effects. The shape of quasi-soliton can be written in the form $\Psi(x,y,t) = \int_{-K/2}^{+K/2} A(\kappa,t)\mathcal{U}(x,y,k+\kappa)e^{i\varepsilon t+i(k+\kappa)y}d\kappa$, where spinor $\mathcal{U} = (u_+, u_-)^T$ satisfies the linear equation $(\mathcal{L}+\varepsilon)\mathcal{U}e^{iky} = 0$ and therefore describes spatial shape of the linear Bloch mode with momentum $k$, and we also take into account that corresponding energy $\varepsilon$ also depends on quasi-momentum $k$. Here $\kappa$ is the momentum offset from the carrier soliton momentum $k$ and the amplitude $A(\kappa,t)$ is assumed well localized in $\kappa$. Using Taylor series expansion in $\kappa$ for $\mathcal{U}(x,y,k+\kappa)$ in the above integral one arrives at the expression for the shape of the edge state wavepacket:

$$\Psi(x,y,t) = e^{i\varepsilon t+iky} \sum_{n=0,\infty} \frac{(-i)^n}{n!} (\partial^n \mathcal{U}/\partial k^n)[\partial^n A(y,t)/\partial y^n], \quad (5)$$

where $A(y,t) = \int_{-K/2}^{+K/2} A(\kappa,t)e^{i\kappa y}d\kappa$ is the envelope function of the corresponding nonlinear edge state. To calculate how $\mathcal{L}$ acts on the spinor wavefunction $\Psi$ we move $\mathcal{L}$ through the integral in the expression for $\Psi$ and take into account that $\mathcal{L}\mathcal{U}e^{iky} = -\varepsilon\mathcal{U}e^{iky}$: $\mathcal{L}\Psi = -\int_{-K/2}^{K/2} \varepsilon(k+\kappa)A(\kappa,t)\mathcal{U}(x,y,k+\kappa)e^{i\varepsilon t+i(k+\kappa)y}d\kappa$. Using Taylor series expansion in $\kappa$ for both $\varepsilon(k+\kappa)$ and $\mathcal{U}(x,y,k+\kappa)$ by analogy with (5) one obtains:

$$\mathcal{L}\Psi = -e^{i\varepsilon t+iky} \sum_{n=0,\infty} \frac{(-i)^n}{n!}[\partial^n (\varepsilon\mathcal{U})/\partial k^n][\partial^n A(y,t)/\partial y^n], \quad (6)$$

We further assume that the spinor $\mathcal{U}$ changes with $k$ much slower than the eigenvalue $\varepsilon$, that is a valid assumption as simulations show. This allows to keep only $n=0$ term in (5), so that $\Psi(x,y,t) = e^{i\varepsilon t+iky}\mathcal{U}(x,y,k)A(y,t)$ and write $\partial^n(\varepsilon\mathcal{U})/\partial k^n \approx \mathcal{U}\partial^n \varepsilon/\partial k^n$ in Eq. (6). The nonlinear term $\mathcal{N}\Psi$ in Eq. (2) acquires in this case a particularly simple form $A|A|^2 \mathcal{N}\mathcal{U}e^{i\varepsilon t+iky}$. Finally, we multiply the equation $i\partial \Psi/\partial t = \mathcal{L}\Psi + \mathcal{N}\Psi$ by $\mathcal{U}^\dagger$ from the left and integrate it over one period along the $y$-axis and along the entire $x$-axis, assuming slow variation of the envelope function $A(y,t)$ with $y$, that allows to remove it from all integrands. This yields the nonlinear Schrödinger equation for the envelope function:

$$i\frac{\partial A}{\partial t} = i\varepsilon'\frac{\partial A}{\partial y} + \frac{1}{2}\varepsilon''\frac{\partial^2 A}{\partial y^2} + gA|A|^2, \quad (7)$$

where we kept only first two terms proportional to $\varepsilon' = \partial\varepsilon/\partial k$ and $\varepsilon'' = \partial^2\varepsilon/\partial k^2$ in the Taylor expansion for $\varepsilon(k)$. The effective nonlinear coefficient is given by $g = \iint \mathcal{U}^\dagger \mathcal{N}\mathcal{U} dxdy / \iint \mathcal{U}^\dagger \mathcal{U} dxdy$. This coefficient can be calculated numerically for different $k$ values using shapes of linear Bloch modes. It turns out to be always positive. When $g\varepsilon'' > 0$ (note, that $\varepsilon'' > 0$ corresponds in our notations to negative effective polariton mass) Eq. (7) admits bright soliton solutions

$$A(y,t) = [2(\varepsilon-\mu)/g]^{1/2} \text{sech}\{[2(\varepsilon-\mu)/\varepsilon'']^{1/2}(y+\varepsilon't)\} \times \exp[i(\mu-\varepsilon)t], \quad (8)$$

where $\mu - \varepsilon \leq 0$ is the energy shift due to repulsive nonlinearity. Note that energy shift in (8) is introduced such that total wavefunction $\Psi(x,y,t) = e^{i\varepsilon t+iky}\mathcal{U}(x,y,k)A(y,t)$ varies in time as $e^{i\mu t}$. The expression (8) is valid as long as the envelope function is much wider than the $y$-period $3^{1/2}a$ of the array.

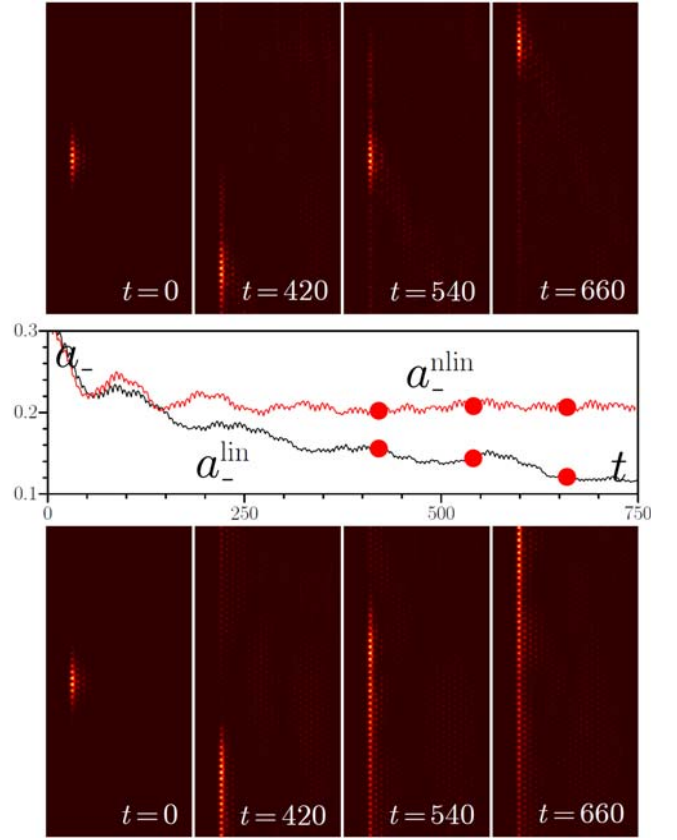

Fig. 7. (Color online) Comparison of nonlinear (top row) and linear (bottom row) evolution of edge state with localized soliton envelope at $\mu - \varepsilon = -0.02$, $k = 0.55\,\text{K}$. Middle row shows peak amplitude $a_-$ of $\psi_-$ component for linear and nonlinear cases. $|\psi_-|$ distributions in the top and bottom rows correspond to red circles. Notice that edge states in all panels in this figure move upwards. To simultaneously stress the fact that edge state moves in one direction and to provide details of its shape, we show $\psi_-$ distributions within relatively small vertical window, but we also applied identical vertical shift for distributions at time moments $t = 420, 540, 660$, where edge state in nonlinear medium remains nearly invariable. The distribution at $t=0$ was not shifted. Here $\beta = 0.3$, $\Omega = 0.5$.

Figure 3(b) shows dispersion coefficient $\varepsilon''$ as a function of $k$ for different linear edge states: we use the same colors as in Fig. 2(c) to denote different branches. One can see that there exist momentum intervals where dispersion coefficient $\varepsilon''$ is positive (effective polariton mass is negative). All such intervals for every linear edge mode give rise to unidirectional edge quasi-solitons. Note, that for every branch there exist a unique $k$ value where dispersion $\varepsilon''$ vanishes and where excitation with broad envelope may evolve almost without broadening even in the linear limit [shape distortions in this case will be determined by weaker $\varepsilon'''$ dispersion that was omitted in (7)].

Top row of Fig. 7 shows the central result of this work – evolution of the quasi-soliton constructed using Eq. (5) with the envelope function given by Eq. (8) in the nonlinear topological insulator state. This quasi-soliton corresponds to the red branch in Fig. 2(c) and $k = 0.55\,\text{K}$. This $k$ value was selected approximately in the *middle* of the domain (in terms of $k$) where dispersion coefficient $\varepsilon''$ for linear edge modes from red branch of Fig. 2(c) is positive. Indeed, this branch exists at $0.24 < k/\text{K} < 0.62$, while dispersion changes its sign at $k \approx 0.44\,\text{K}$.

The particular momentum value $k = 0.55\,\text{K}$ at which we demonstrate edge soliton formation is not preferred in comparison with other $k$ values at which $\varepsilon'' > 0$: we were able to generate similar long-living nonlinear edge states for multiple values of $k$.

One can see that after the initial transient where the peak amplitude decreases due to internal reshaping of the input, the unidirectional quasi-soliton forms whose amplitude remains almost constant in time and whose velocity nearly coincides with $\varepsilon'$ (see red curve in the central row with dependence of peak amplitude $a_-^{\text{nlin}}$ of $\psi_-$ component). Note, that, during time period shown, the quasi-soliton traverses over 100 periods of the array. One can observe small radiation into the depth of array that gradually reduces soliton amplitude (that is why we call such states quasi-solitons), but even on the time scales $t \sim 10^3$ it is a negligible effect, so that quasi-solitons found here are exceptionally robust at any time scales. Similar unidirectional states were found for other dispersion branches (thus counterparts of soliton from Fig. 7, but from green branch move in the opposite $y$-direction) and also in arrays with bearded edges.

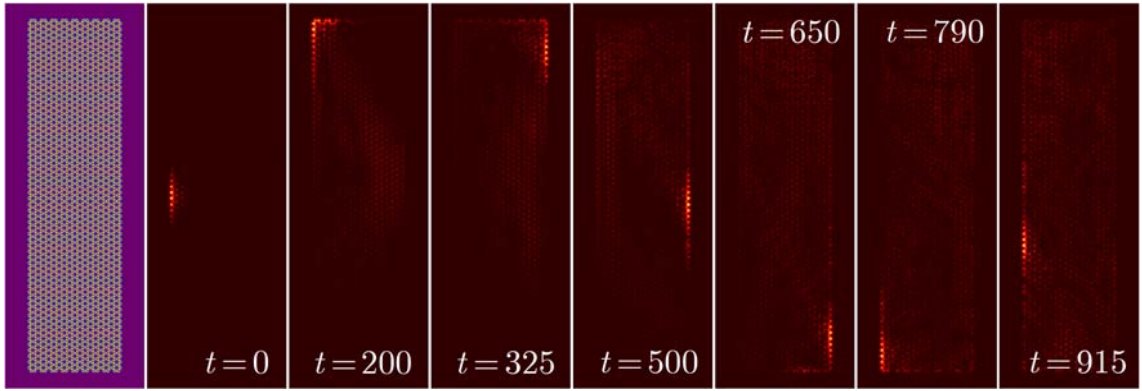

Fig. 8. (Color online) Evolution of edge state with localized soliton envelope at $\mu - \varepsilon = -0.02$, $k = 0.55\,\text{K}$ in rectangular array. Only $|\psi_-|$ distributions are shown. In all cases $\beta = 0.3$, $\Omega = 0.5$. Left plot shows array.

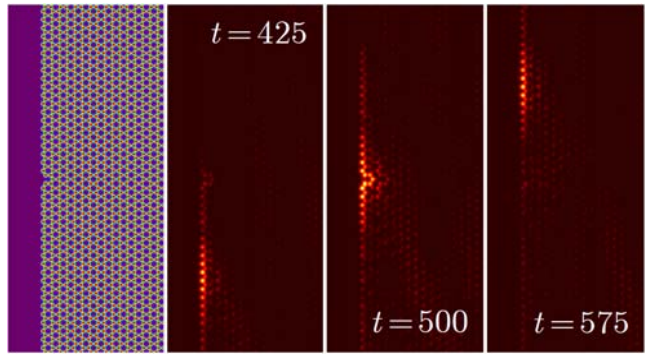

Fig. 9. (Color online) Passage of soliton obtained from edge state with localized envelope at $\mu - \varepsilon = -0.02$, $k = 0.55\,\text{K}$ through surface defect. Soliton was allowed to move along the surface of array over considerable time interval before collision with defect. Only $\psi_-$ component is shown. Here $\beta = 0.3$, $\Omega = 0.5$.

To confirm that the obtained localized states indeed exist due to nonlinear self-action, we used the same input, but switched off nonlinearity, see the bottom row in Fig. 7. Without nonlinearity we have observed a pronounced asymmetric expansion of the wavepacket, while the peak amplitude $a_-^{\text{lin}}$ of $\psi_-$ component was strongly decreasing upon evolution (see black curve in the middle row). We have also checked that linear edge states with $\varepsilon'' < 0$ do not give rise to quasi-solitons.

To confirm exceptional robustness of quasi-solitons and their immunity to backscattering in nonlinear regime we considered their evolution in the lattice potential that was made finite also along the $y$-axis (Fig. 8). We used the same quasi-soliton input as in Fig. 7. One can see that the soliton survives even upon passage of several array corners and that it returns to its initial location after making a closed loop along the surface of the array with minimal decrease in peak amplitude of both components. Note, that the radiation into the depth of array is pronounced only along the top and bottom armchair edges that formally do not support edge states and that is why they were made relatively short.

Finally, the absence of backscattering on edge defects is illustrated in Fig. 9, where we removed one of the micropillars on the surface of the array. Before collision with this defect, the quasi-soliton was allowed to evolve over sufficiently long time to ensure that its amplitude has reached a "steady state" value. Note, that the soliton experiences considerable reshaping at the point of collision with the defect, so that its amplitude drops nearly by a factor of 2, but it returns to its initial value immediately after passage of the defect.

## 5. Summary

Summarizing, we predicted formation of extended nonlinear edge states in polariton condensates with spin-orbit coupling held in the honeycomb lattice potentials. Nonlinear edge states spontaneously decay into sets of fully localized quasi-solitons that can also be selectively excited by using proper envelope function, derived in this paper. The edge solitons have been shown to be robust with respect to the passage through the intervals of the lattice edges that do not support the edge states in the linear approximation and through other lattice defects.

### Acknowledgements

We gratefully acknowledge support from Russian Federal Target Program (RFMEFI58715X0020); European Union Seventh Framework Programme (FP7) (LIMACONA 612600); Leverhulme Trust (RPG-2012-481); Severo Ochoa Excellence program of the government of Spain.